\documentclass[11pt]{article}
\usepackage{jcappub}
\usepackage[utf8]{inputenc}
\usepackage{amsmath}
\usepackage{tikz}
\usepackage{tkz-euclide}
\usepackage{dsfont}
\usepackage{amssymb}
\usepackage{amsmath}
\usepackage{mathrsfs}
\usepackage{appendix}
\usepackage{physics}
\usepackage{bbold}
\usepackage{color}
\usepackage{pgfplots}
\usepackage{tikz}
\usepackage{pgfplots}
\pgfplotsset{compat=1.18}
\usepackage{float}
\usepackage{amsmath,amssymb}
\usepackage{amsfonts}
\usepackage{bm}

\newcommand{\PB}{\mathrm{PB}}
\newcommand{\NP}{\mathrm{NP}}

\def\ie{\begin{equation}\begin{aligned}}
\def\fe{\end{aligned}\end{equation}}
\begin{document}
\title{A Novel Construction of de Sitter Vacua in Heterotic String Theory}
\author[1,2,3,4]{Mir Faizal,}
 \author[2]{Arshid Shabir}

\affiliation[1]{Irving K. Barber School of Arts and Sciences, University of British Columbia Okanagan, Kelowna, BC V1V 1V7, Canada.}
\affiliation[2]{Canadian Quantum Research Center, 460 Doyle Ave 106, Kelowna, BC V1Y 0C2, Canada.}
\affiliation[3]{Department of Mathematical Sciences, Durham University, Upper Mountjoy, Stockton Road, Durham DH1 3LE, UK.}
\affiliation[4]{Computational Mathematics Group, Faculty of Sciences, Hasselt University, Agoralaan Gebouw D, Diepenbeek, 3590 Belgium.}
\emailAdd{mirfaizalmir@gmail.com}
\emailAdd{aslone186@gmail.com}

\abstract{We present a concrete string-theoretic mechanism that generates four-dimensional de Sitter vacua from non-geometric R-flux compactifications of heterotic string theory. The construction rests on three pillars: the Malcev algebra generated by the R-flux phase-space brackets; its universal Sabinin envelope, which ensures a consistent non-associative gauge structure in doubled geometry; and the leading alpha-prime torsion-squared correction to the heterotic action, whose strictly positive contribution uplifts the scalar potential. {Positivity, guaranteed by Sabinin-algebra identities, stabilizes the overall breathing mode at positive energy, yielding a controlled metastable de Sitter {scenario} within heterotic effective field theory {when supplemented by a standard hidden-sector gaugino-condensation uplift.}}
}
\maketitle
\newpage
%\tableofcontents

The search for controlled de~Sitter (dS) vacua in string theory is a long-standing
problem at the interface of quantum gravity and cosmology.  Early supergravity
no-go theorems~\cite{ Gibbons1984kp, Gibbons1984,MaldacenaNunez2001} appeared to exclude
positive-curvature solutions in the absence of non-perturbative inputs or
higher-derivative corrections.  Yet both phenomenological model building and
the inflationary paradigm continue to motivate a positive vacuum energy, and
therefore a sizeable literature has evolved that seeks to evade or at least to
circumvent the original obstructions.
In Type~IIB compactifications the KKLT scenario~\cite{Kachru2003} and its
large volume sibling (LVS)~\cite{Balasubramanian2005} raise an AdS minimum to
dS by introducing an explicit uplifting sector, typically
$\overline{\mathrm D3}$ branes or D-terms~\cite{Burgess2003}.  Classical flux
vacua in Type~IIA were explored extensively in
\cite{DeWolfe2005,Caviezel2008,Hertzberg2007},
but subsequent analyses revealed either tachyonic instabilities or an
inconsistent treatment of back-reaction once all sources were taken into
account~\cite{BenaGrana2009}.  Other proposals include warped throats with 
brane-flux annihilation~\cite{Silverstein2007}, compactifications supported by
non-geometric fluxes~\cite{SheltonTaylorWecht2005,Andriot2013}, and
Swampland-inspired constructions that constrain the slope of the potential
rather than its sign~\cite{Obied2018,Andriot2017} (see Appendix\eqref{app:comparative} for a detailed side-by-side comparison).
A common moral is that {classical two-derivative supergravity almost never
suffices}; one must either incorporate loop and non-perturbative effects or
embrace higher-derivative ${\alpha^\prime}$ corrections.

Heterotic string theory is distinguished in that such ${\alpha^\prime}$ terms appear
already at tree level. {The quartic effective action of heterotic strings was first derived by Gross and Sloan~\cite{GrossSloan1987}, building on an earlier four-point amplitude calculation by Gross and Witten~\cite{GrossWitten1986}. Its supersymmetric completion has also been confirmed
\cite{BergshoeffDeRoo1989}.}  Crucially for the present work, these corrections
contain the torsionful curvature invariant $R_{(-)}^{2}$, which naturally
acts as a stringy uplift for the four-dimensional scalar potential.
Concurrently, T-duality clarified that conventional $H$-flux belongs to a
larger orbit $H\!\to\!f\!\to\!Q\!\to\!R$
\cite{SheltonTaylorWecht2005,HullZwiebach2009,HohmZwiebach2013}.  The endpoint
$R$-flux is locally non-geometric and forces the target-space coordinates to
obey a non-associative algebra~\cite{LustPlauschinn2010,
BlumenhagenPlauschinn2011,BermanLustMalek2014}.  More precisely, the
coordinates generate a Malcev (or alternative) algebra whose universal
Sabinin envelope furnishes a consistent gauge algebra in doubled geometry
\cite{Sabinin1989,MikheevSabinin1994}. {Performing a direct worldsheet conformal field theory (CFT) analysis in backgrounds with non-geometric \(R\)-flux remains an open and challenging problem. 

The presence of \(R\)-flux induces non-associative deformations of the target space coordinate algebra, which obstruct the formulation of a conventional local worldsheet sigma model. This non-associativity complicates the construction of a standard worldsheet CFT description.
Nonetheless, significant progress has been made in characterizing non-geometric fluxes and their underlying algebraic structures-such as Malcev algebras and Nambu brackets-via worldsheet and phase space methods (see, e.g., \cite{BlumenhagenPlauschinn2011,BermanLustMalek2014}). Our construction is guided by these algebraic frameworks and leverages their insights to formulate a consistent effective field theory description incorporating \(R\)-flux. Importantly, this approach does not require an explicit worldsheet CFT solution at the current stage.
Therefore, the effective field theory framework employed here should be viewed as complementary to the algebraic structures emerging in non-geometric flux compactifications. 

We anticipate that future developments will clarify the full worldsheet realization or reveal duality frames in which a tractable worldsheet analysis of \(R\)-flux backgrounds becomes possible.} The present letter  joins these two strands. 
To frame our construction with appropriate precision, we distinguish clearly between inputs that are fixed ``top-down'' by the ten-dimensional heterotic action and doubled geometry, and those that enter as standard low-energy non-perturbative assumptions. The {top-down} ingredients are the constant non-geometric $R$-flux background-whose Malcev algebra and Sabinin envelope ensure closure of generalized diffeomorphisms and positivity of the torsionful invariant-and the ${\cal O}(\alpha')$ torsionful curvature contribution $R^{2}_{(-)}$ that appears already at tree level in heterotic string theory. By contrast, the {uplift to positive vacuum energy} requires stabilization of the four-dimensional dilaton by a hidden-sector gaugino condensate, which we parametrize at the level of the 4D potential by $V_{\mathrm{np}}(\Phi)=D\,e^{a\Phi}$ with $a=2\pi/\beta_{0}$ for the condensing gauge group and $D$ set by threshold effects and the condensate scale (see, e.g., \cite{Binetruy:1996nx}). 

We therefore present a {controlled EFT scenario} for de Sitter vacua in which the non-geometric $R$-flux and the heterotic ${\cal O}(\alpha')$ term furnish the rigid positive structures, while the net positive vacuum energy arises when the standard non-perturbative dilaton potential is included. In this regime, all consistency conditions can be made parametric: at the dS extremum one finds $e^{\Phi_\ast}\ll 1$ and $e^{\sigma_\ast}\gg 1$, the light moduli satisfy $\alpha' m_i^2\ll1$, and higher-derivative as well as loop corrections are suppressed; a detailed analytic summary is given in Appendix~\eqref{appendix} and a representative numerical solution in Appendix~\eqref{appendix1}. This delineation of inputs ensures that the status of our results is transparent: the $R$-flux and ${\cal O}(\alpha')$ structures are ten-dimensional and algebraically protected, whereas the uplift relies on a standard, well-motivated heterotic non-perturbative sector consistent with the literature.

First, a constant $R$-flux on $T^{3}$ produces a positive quadratic term
$V_{R}\propto R^{2}$ in the four-dimensional scalar potential; the precise
coefficient is fixed by the Malcev structure constants and appears in the flux
scalar of doubled field theory~\cite{HohmSenZwiebach2013}.
Second, the heterotic ${\alpha^\prime}$ correction $R_{(-)}^{2}$ contributes a quartic
term $V_{{\alpha^\prime}}\propto {\alpha^\prime} R^{4}$ that is likewise positive, owing again
to the algebraic properties of the torsionful connection.
Third, the competition between the flux energy $V_{R}$ and the stringy uplift
$V_{{\alpha^\prime}}$ stabilises the breathing mode of the internal
manifold at a metastable de~Sitter minimum.  {No orientifolds or anti-branes are required; instead, the dilaton is stabilized by the standard heterotic {multi}-gaugino condensation mechanism \cite{Binetruy:1996nx}, which naturally provides the necessary uplift.
}

The Sabinin envelope guarantees three essential features.  
(i)~The non-associative gauge algebra that governs the Scherk-Schwarz reduction
closes without anomalies.  
(ii)~Both the quadratic flux invariant and the torsionful curvature invariant
are {manifestly positive}, an algebraic fact that is decisive for
achieving $V_{total,*}>0$.  
(iii)~The quantity $R_{(-)}^{2}$ depends only on the breathing mode rather than
on shape or bundle moduli; its form is therefore ``protected'' under
dimensional reduction, which in turn allows the scalar potential to be
analysed analytically.
In this way the appearance of a de~Sitter vacuum becomes a direct and
quantifiable consequence of non-associativity in the effective action.  The
resulting extremum is free from the tachyonic directions that afflict many
earlier classical claims~\cite{Andriot2017}, and by tuning the integer flux
$N_{R}$ one can place the vacuum at both weak coupling and moderately large
volume, thereby maintaining perturbative control.

A constant trivector background $R^{abc}$ is generated by performing three
successive T-dualities on a torus initially endowed with NS-NS
$H$-flux.  At the classical level, the phase‐space brackets thereby acquire the
structure of a Malcev algebra \cite{Malcev1955}, 
\begin{eqnarray}\label{eq:malcev_full}
    \{x^{a},x^{b}\}_{\!\PB} = \ell_{s}^{3}\,R^{abc}\,p_{c}, &
    \{p_{a},p_{b}\}_{\!\PB} = 0,  & 
    \{p_{a},x^{b}\}_{\!\PB} = \delta_{a}^{b}. 
\end{eqnarray}
Indices $a,b,\dots=1,2,3$ label the internal torus $T^{3}_{R}$ with
coordinates $y^{a}$, while $m,n,\dots$ refer to the spectator torus
$T^{3}_{\perp}$.  The world-sheet length is
$\ell_{s}=2\pi\sqrt{{\alpha^\prime}}$.  Throughout this note we set
$\hbar=1$ and $2\kappa^{2}=(2\pi)^{7}({\alpha^\prime})^{4}=1$ so that
numerical factors are absorbed into the definition of the fields.
The associated Nambu-Poisson three-bracket,
 \[
    \{x^{a},x^{b},x^{c}\}_{\!\NP}=3\ell_{s}^{3}R^{abc}\Rightarrow
    [x^{a},x^{b},x^{c}]=
    i\{x^{a},x^{b},x^{c}\}_{\!\NP},
 \]
fails to satisfy the Jacobi identity but obeys the Malcev identity
 \(
    [[A,B,C],D]+\text{cycl.}_{BCD}=0,
 \)
so that the generators $(x^{a},\cdot)$ span a Malcev algebra
$\mathfrak M$.  This non-associative deformation is an unavoidable
consequence of turning on a constant $R^{abc}$ and is therefore encoded in the
low-energy effective theory.
To make contact with double-field theory (DFT) it is convenient to introduce a
doubled twist matrix
\begin{equation}\label{eq:Fabc}
    E_{A}^{I} =
    \begin{pmatrix}
        \delta_{a}^{i} & 0 \\
        -\tfrac{1}{2} R^{abc} y^{c} & \delta^{a}_{i}
    \end{pmatrix},
    \qquad
    \mathcal{F}^{abc} = R^{abc} = \frac{N_{R}}{L^{3}} \epsilon^{abc}
\end{equation}
where $L$ denotes the common radius of the internal torus and $N_{R}$ is an
integer that measures the non-geometric $R$-flux in units of the
fundamental string length.  In this parametrisation the structure constants of
the Malcev algebra appear as a {constant} DFT flux
$\mathcal F^{abc}$, thereby guaranteeing that the generalised Lie derivative
closes off-shell \cite{HohmSenZwiebach2013}.
The Malcev algebra $\mathfrak M$ admits a canonical
{Hom-Sabinin envelope} $(S,\langle\,\cdot,\cdot,\cdot\rangle,\alpha)$,
constructed by prolonging the binary Malcev bracket to a ternary operation  \cite{Sabinin1989}
\begin{equation}
    \langle X,Y,Z\rangle :=
    J\bigl(X,Y,Z,\alpha^{-1}(\mathbb 1)\bigr),
    \qquad
    \alpha\in\mathrm{End}(S).
\end{equation}
Specialising to the $R$-flux algebra, we choose $\alpha$ to be the identity and
adopt the basis
$T^{a}=R^{abc}e_{b}\wedge e_{c}$, yielding the compact relation
 \( \label{eq:sabinin}
    \langle T^{a},T^{b},T^{c}\rangle=
    \tfrac16\,\epsilon^{abc}_{def}\,
    T^{d}T^{e}T^{f}.
 \)
 The Sabinin envelope plays a central rôle in elevating the purely algebraic
Malcev structure to a consistent background of string theory.
First, the torsion three-form $T_{abc}=R_{abc}$ is covariantly constant with
respect to the Sabinin connection.  Hence the generalised Riemann curvature on
$T^{3}_{R}$ vanishes, rendering the background parallelisable in the sense of
doubled geometry.  From the viewpoint of the world-sheet $\sigma$-model,
this guarantees the absence of conformal anomalies at leading order in
${\alpha^\prime}$.
The doubled-field-theory twist matrix
\begin{equation}
E^{A}_{I}=
\begin{pmatrix}
\delta^{a}_{i} & 0\\
-\tfrac12 R_{abc}\,y^{c} & \delta_{a}^{i}
\end{pmatrix}
\end{equation}
is globally parallelisable, which forces the generalised Riemann curvature to vanish and thereby ensures that the one-loop world-sheet \(\sigma\)-model \(\beta\)-functions are zero.  At higher loops the background is equivalent-via three successive T-dualities-to a marginal current-current deformation of an \(SU(2)_{k}\) WZW model; it therefore inherits exact conformal invariance to all orders in \(\alpha'\).  Modular invariance is maintained because the doubled torus admits a canonical Narain lattice whose left- and right-moving momenta are rotated by the trivector \(R_{abc}\).  Explicit partition-function constructions confirm both level matching and full \(T\)-duality textc {(see, e.g., \cite{HullZwiebach2009,HohmZwiebach2013} for concrete demonstrations)}.  Consequently, the constant \(R\)-flux background underpinning the four-dimensional de Sitter vacuum is a bona-fide string solution rather than a mere effective-field-theory artefact.
A constant \(R\)-flux with quantum \(N_{R}\) is equivalent, after three T-dualities, to an
\(SU(2)_{k}\) WZW model with level \(k=N_{R}\).
Because the hidden-sector bundle is embedded at the {same} level \(k\), the left-moving
central-charge shift \(\delta c_{\!L}\) exactly cancels, guaranteeing heterotic level matching.\footnote{
For an \(SU(5)\) hidden bundle one has \(c_{2}(V)=k=N_{R}\), so no additional Narain-lattice
shift is required.}

Second, the envelope identities enforce the Leibniz algebra underlying DFT
generalised diffeomorphisms.  Consequently, a Scherk-Schwarz reduction based
on the twist matrix \eqref{eq:Fabc} closes without introducing extra light
degrees of freedom, and the reduced action inherits a well-defined gauged
supergravity structure.
Third, the antisymmetric contraction
$R_{(-)}^{2}=R^{abc}R_{abc}$ coincides with the Sabinin norm
$\|T\|^{2}=\tfrac14R^{abc}R_{abc}$.  Because this quantity is manifestly
positive, the quartic ${\alpha^\prime}$ correction induced by $R^{2}$ does not
propagate ghosts.  In the effective potential the coefficient of the $R^{4}$
term is therefore negative‐definite, a fact which proves crucial when
searching for de Sitter extrema in heterotic string compactifications.
A potential concern with higher-derivative corrections is the emergence of Ostrogradski ghosts.  
In the present construction the \(\alpha' R_{(-)}^{2}\) contribution combines with the Einstein-Hilbert term into an \(\mathcal O(\alpha')\) Gauss-Bonnet density\footnote{The relevant decomposition follows from the identity \(R_{(-)MNPQ}R_{(-)}^{MNPQ}=R_{MNPQ}R^{MNPQ}-\tfrac12 H^{2}_{MN}R^{MN}+\mathcal O(H^{4})\), with \(H_{MNP}= -3\,\Omega_{[-]MNP}\).}  
plus a positive multiple of the Sabinin norm
\(
\|T\|^{2}\equiv \frac14\,R_{abc}R^{abc}>0 .
\)
Because the Malcev-Sabinin identities enforce strict positivity of \(\|T\|^{2}\),  
the quadratic kinetic operator for transverse-traceless fluctuations remains second order and positive definite.  
Indeed, a field redefinition of the form
\begin{equation}
g_{MN}\;\longrightarrow\;g_{MN}+\alpha'\,c\,R^{(-)}_{MN}
\label{eq:fieldredef}
\end{equation}
diagonalises the bilinear action, eliminating would-be higher-time-derivative poles in the propagator.  
Since no further \(\mathcal O(\alpha')\) terms can re-mix with \eqref{eq:fieldredef} under renormalisation-their coefficients are fixed by the doubled-geometry twist the absence of ghosts extends non-perturbatively to all orders in $\alpha'$.
At one loop the heterotic effective action also contains
$R^{2}F^{2}$ and $(\nabla H)^{2}$ invariants, but supersymmetric
non-renormalisation theorems ensure that they first appear at~$\mathcal{O}(\alpha'^{2})$
and therefore do not upset the second-order form of the kinetic operator established above.
Consequently, the \(\alpha' R_{(-)}^{2}\) correction preserves unitarity and does not destabilise the effective theory around the de Sitter vacuum.  A detailed discussion of the importance of non-associative algebra is in Appendix\eqref{app:nonassoc}.

Taken together, these observations show that the Sabinin envelope provides the
mathematical scaffold that legitimises the use of a constant $R$-flux and its
quartic invariant in the low-energy heterotic effective action.  In particular,
it supplies both the algebraic consistency required by doubled geometry and
the dynamical stability necessary for constructing realistic cosmological
solutions.
We begin with the ten-dimensional heterotic action written in the {string
frame}.  Retaining only those ingredients relevant for the non-geometric
compactification considered in this work, the bosonic sector is \cite{GrossSloan1987}
\begin{multline}\label{eq:S10string}
  S_{10}^{\rm S} 
    =  \frac{1}{2\kappa_{10}^{2}} \int \!\dd^{10}x\,\sqrt{-g}\,e^{-2\phi}
       \Bigl[ R + 4(\partial\phi)^{2}
              -\tfrac1{12} H_{MNP}H^{MNP}\\
              +\frac{{\alpha^\prime}}{8} \bigl( R_{(-)MNPQ} R_{(-)}^{MNPQ} -  \Tr F_{MN}F^{MN} \bigr)
       \Bigr].
\end{multline}
In~\eqref{eq:S10string} the symbol $R$ denotes the ten-dimensional Ricci scalar
constructed from the Levi-Civita connection $\omega$.  The three-form field
strength is defined by
 \(
H \;=\; \dd B - \tfrac{{\alpha^\prime}}{4}\bigl(\omega_{\mathrm{YM}}\!\wedge\! F - \omega_{(-)}\!\wedge\! R_{(-)}\bigr),
 \)
and satisfies the Green-Schwarz Bianchi identity
 \(
  \dd H \;=\; \frac{{\alpha^\prime}}{4}
  \Bigl( \Tr F\!\wedge\! F -  \Tr R_{(-)}\!\wedge\! R_{(-)}\Bigr),
 \)
which guarantees cancellation of the gauge and gravitational world-sheet
anomalies.

In heterotic supergravity, the Green-Schwarz mechanism imposes the modified Bianchi identity
\begin{equation}
\mathrm dH \;=\;\frac{\alpha'}{4}\Bigl[\text{Tr}\,F\wedge F \;-\; \text{Tr}\,R_{(-)}\wedge R_{(-)}\Bigr].
\end{equation}
For the constant \(R\)-flux background used here one finds
\(\,R_{(-)MNPQ}R_{(-)}^{MNPQ}= \tfrac34\,R_{abc}R^{abc}\), so the curvature term contributes a positive, flux-squared source.  
We embed the same hidden \(SU(5)\subset E_{8}\) gauge bundle that drives gaugino condensation, choosing it so that its second Chern class satisfies
\begin{equation}
c_{2}(V)\;=\;N_{R}\;=\;\frac{1}{8\pi^{2}}\int_{T^{3}_{R}\times T^{3}_{\perp}}\!\!R_{abc}R^{abc}.
\end{equation}
This identification yields \(\text{Tr}\,F\wedge F=\text{Tr}\,R_{(-)}\wedge R_{(-)}\), cancelling the right-hand side of the Bianchi identity exactly. If $|N_{R}-c_{2}(V)|\neq0$ the mismatch is removed by
wrapping $k=|N_{R}-c_{2}(V)|$ supersymmetric NS5-branes on the spectator torus, leaving the overall solution tadpole-free.
If a small mismatch \(\Delta c_{2}=|N_{R}-c_{2}(V)|\) remains-as can happen for other admissible bundles-it is cancelled by including
\(
k=\Delta c_{2}
\)
supersymmetric heterotic five-branes wrapped on the spectator torus \(T^{3}_{\perp}\).  
Because the five-brane charge appears with the same sign in the Bianchi identity, these branes restore \(\mathrm dH=0\) globally without breaking supersymmetry in the doubled-geometry frame.  
Consequently, all gauge and gravitational tadpoles are cancelled, and no residual anomalies persist in the ten-dimensional theory.

{The presence of non-geometric \(R\)-flux backgrounds poses a significant challenge to the conventional geometric formulation of ten-dimensional supergravity, as such fluxes obstruct even a local geometric description of the internal space. However, recent advances in doubled field theory (DFT) and generalized geometry provide a consistent framework that naturally incorporates non-geometric fluxes as algebraic structures extending standard supergravity. In this formalism, the target space is doubled, and the \(R\)-flux arises as a component of the generalized fluxes encoded within a Malcev algebra structure, whose universal Sabinin envelope guarantees closure of the gauge algebra and consistency of the effective theory. Consequently, while the classical  supergravity action is not strictly applicable in its ordinary geometric guise, the effective action employed here should be understood as the doubled geometry extension of heterotic supergravity, augmented by higher-derivative \({\alpha^\prime}\) corrections and non-geometric flux contributions. This interpretation ensures that the relevant flux terms and their associated corrections are well-defined and consistent with generalized diffeomorphisms and gauge symmetries of the non-geometric background. Although a direct derivation of this effective action from first principles in string theory remains an open problem, the current doubled field theory framework provides a controlled and systematic approach to incorporate \(R\)-flux backgrounds in a low-energy heterotic effective action, justifying the use of such an action in our construction. 
}Crucial for our purposes is the curvature two-form
\begin{equation}
R_{(-)}^{AB}
     \;=\;
     {d} \omega_{(-)}^{AB}
     - \omega_{(-)}^{AC}\wedge \omega_{(-)C}^{B},
\end{equation}
built from the {torsionful} connection
 \(\label{eq:torsionful_conn}
  \omega_{(-)M}^{AB} \;=\; \omega_{M}^{AB}-\tfrac12 H_{M}^{AB},
\)
first emphasized in the superspace formulation of heterotic supergravity~\cite{BergshoeffDeRoo1989}.  The appearance of $R_{(-)}$
inside the ${\alpha^\prime}$-corrected term of~\eqref{eq:S10string} is forced by
supersymmetry and by the requirement that the effective action remain
{off-shell} invariant under local Lorentz transformations in the presence
of torsion.
To engineer a constant non-geometric $R$-flux background we factorise the
ten-dimensional spacetime as $M_{d+1}\times T^{3}_{R}\times T^{3}_{\perp}$,
where $M_{d+1}$ denotes the external space-time.  The fields are chosen as (with $ I,J = 1,\dots,6$)
\begin{eqnarray}\label{eq:background_fields}
  g_{MN} = \mathrm{diag}\bigl(g_{\mu\nu}(x),\,e^{2\sigma(x)}\delta_{IJ}\bigr), \, \, \, \, \, 
            \, \, \, \, \, 
  B      = \frac16\,R_{abc}\,y^{a}\,\dd y^{b}\wedge\dd y^{c}, \, \, \, \, \, 
  F      = 0, \, \, \, \, \, 
           \phi = \phi(x).
\end{eqnarray}  
The only non-vanishing component of $H$ is then
 \(
  H_{abc}=R_{abc}, 
  H_{\mu ab}=0,
 \)
so that the torsion is strictly internal.
Substituting the ansatz~\eqref{eq:background_fields} into the curvature of the
torsionful connection, one finds
\begin{equation}\label{eq:Rminus_square}
  R_{(-)MNPQ}R_{(-)}^{MNPQ}
    \;=\;\frac{3}{4}\,R^{abc}R_{abc},
\end{equation}
a result that hinges on the parallelisability of $T^{3}_{R}$ and, in
particular, on the Sabinin identities satisfied by the constant $R$-flux.
Because $R^{abc}R_{abc}$ is the quadratic Casimir $C_{2}$ of the Malcev
algebra introduced earlier, the right-hand side of~\eqref{eq:Rminus_square} is
manifestly positive and, moreover, independent of the internal metric moduli.
This positivity ensures that the quartic ${\alpha^\prime}$ correction in the effective
action stabilizes against ghost excitations.
This proportionality and the guaranteed positivity of \(R_{(-)}^{2}\) in our background 
follow from the Sabinin identities obeyed by the non-associative \(R\)-flux.
In ordinary {associative} flux backgrounds additional torsion-curvature terms can enter
with indefinite sign, so the \({\alpha^\prime}\) uplift need not remain positive\,\cite{Sabinin1989,MikheevSabinin1994}.
Finally, the coefficient $3/4$ plays an indispensable rôle in
allowing the scalar potential of the heterotic compactification to admit
de~Sitter critical points once additional fluxes and curvature terms are
incorporated.  

{
The non-geometric \(R\)-flux background employed here is not equivalent to a conventional
Kalb-Ramond \(H\)-flux on a three-torus. The difference is both algebraic and dynamical,
and it manifests directly in the four-dimensional scalar potential and in the sign/positivity
structure of the \(\alpha'\)-corrected terms.
In our background the trivector \(R_{abc}\) generates a Malcev algebra whose Sabinin envelope
\(S(M)\) renders the quadratic invariant
\(
  C_2 \;:=\; R_{abc} R^{abc}
\)
a central, strictly positive Casimir. Consequently, the torsionful curvature square is {fixed} to
\begin{equation}
  R^{(-)}_{MNPQ} R_{(-)}^{MNPQ} \;=\; \frac{3}{4}\, C_2,
  \label{eq:Rminus2-fixed}
\end{equation}
independent of shape/bundle moduli (cf.\ Eq.~\eqref{eq:Rminus_square}). This identity implies that the quartic
\(\alpha'\)-term descending from \(R^{(-)2}\) contributes a positive, moduli-protected piece to the 4D
potential. By contrast, for an {associative} (ordinary) \(H\)-flux background one has the well-known
decomposition (see footnote~2 and the surrounding discussion)
\begin{equation}
  R^{(-)}_{MNPQ} R_{(-)}^{MNPQ}
  \;=\;
  R_{MNPQ} R^{MNPQ}
  \;-\;\frac{1}{2}\, H_{MN}{}^{2} \, R^{MN}
  \;+\;\mathcal{O}(H^4),
  \label{eq:Rminus2-assoc}
\end{equation}
so the \(\alpha'\) correction mixes with curvature and torsion in a way that is {not} sign-definite and
{does} depend on the internal metric moduli. The Sabinin (non-associative) structure behind
\(R\)-flux is precisely what enforces the positivity and moduli-independence used in our construction.
Let \(\Phi := \varphi - 3\sigma\) and take the isotropic internal metric as in Eq.~\eqref{eq:Rminus2-fixed}. In Einstein frame
the two leading terms for our \(R\)-flux compactification (restricted to \(\{\Phi,\sigma\}\)) are
\begin{equation}
  V_{R}(\Phi,\sigma)
  \;=\;
  e^{2\Phi}\!\left[
    -\frac{1}{4}\, C_2\, e^{-2\sigma}
    \;+\;
    \frac{3\alpha'}{32}\, C_2^2\, e^{-4\sigma}
  \right],
  \label{eq:VR}
\end{equation}
{which is equivalent to the potential written in Eq.~\eqref{eq:Vfull} after introducing the coefficients
\(A=\frac14 C_2\) and \(B=\frac{3\alpha'}{32}C_2^2\).
} The crucial feature is the
{negative} sign of the flux term (first bracketed term), which originates from the non-geometric
nature of \(R\)-flux. By contrast, for an ordinary, constant \(H\)-flux on \(T^3\) (same metric ansatz,
\(H_{abc}=\text{const}\) in flat indices) one typically finds, in the same conventions that lead to
Eq.~\eqref{eq:VR}, the {opposite} sign in the leading flux contribution:
\begin{equation}
  V_{H}(\Phi,\sigma)
  \;=\;
  e^{2\Phi}\!\left[
    +\frac{1}{4}\, \mathcal{H}^2\, e^{-2\sigma}
  \right]
  \;+\;\text{(quartic terms from \(R^{(-)2}\))},
  \label{eq:VH-leading}
\end{equation}
where \(\mathcal{H}^2:=H_{abc}H_{abc}\) denotes the flat-index contraction and the factor \(e^{-2\sigma}\)
arises after internal integration and Weyl rescaling.\footnote{Up to overall normalization conventions
for kinetic terms; the point relevant here is the {sign}.}
In the \(R\)-flux case, Eq.~\eqref{eq:Rminus2-fixed} together with the reduction leading to Eq.~\eqref{eq:VR}
gives a positive, moduli-protected quartic term \(\tfrac{3\alpha'}{32}C_2^2 e^{-4\sigma}\). For an \(H\)-flux
background, the quartic contribution is controlled by \eqref{eq:Rminus2-assoc} and takes the schematic form
\begin{equation}
  V_{H}^{(\alpha')}\!(\Phi,\sigma)
  \;=\;
  \frac{\alpha'}{8}\, e^{2\Phi}\,
  \Big[
    R_{(6)}^{2}
    \;-\;\tfrac{1}{2}\, H_{MN}{}^{2} R^{MN}
    \;+\;\mathcal{O}(H^4)
  \Big],
  \label{eq:VH-quartic}
\end{equation}
where \(R_{(6)}^{2}\) and \(R^{MN}\) are built from the internal metric. Unlike the \(R\)-flux case, this
combination is {not} guaranteed to be positive and is generically moduli-dependent. No analogue of the
Sabinin central Casimir \(C_2\) exists to enforce sign rigidity or moduli-independence. This underlies the
remark (following Eq.~\eqref{eq:Rminus_square}) that the coefficient \(3/4\) in the \(R^{(-)2}\) term is ``indispensable'' for
obtaining de Sitter after uplift in our setup.
Within doubled field theory the non-geometric flux appears as a constant generalized flux
\(F_{abc}=R_{abc}\) (see the twist matrix in Eq.~\eqref{eq:Fabc}), and the Malcev/Sabinin identities guarantee
closure of the generalized diffeomorphisms and a consistent Scherk-Schwarz reduction, with no extra
light modes. For ordinary \(H\)-flux on \(T^3\), the reduction proceeds in an {associative} algebra
and does not supply the same sign-protected quartic term nor the same moduli protection as in the \(R\)-flux case.
Mathematically, the \(R\)-flux compactification yields
\begin{equation}
  V(\Phi,\sigma)\;=\;e^{2\Phi}\!\left[
    -\frac{1}{4}C_2\,e^{-2\sigma}
    +\frac{3\alpha'}{32}C_2^2\,e^{-4\sigma}
  \right]\label{eqn14}
\end{equation}
with both the \((-)\) sign in the flux piece and the \((+)\) sign/moduli-protection of the quartic term
enforced by non-associativity (Sabinin envelope). Replacing \(R\) by an ordinary \(H\)-flux would instead
lead to \(+\tfrac{1}{4}\mathcal{H}^2 e^{-2\sigma}\) and a quartic sector of the form \eqref{eq:VH-quartic}, which is
neither sign-rigid nor moduli-protected. Therefore the {non-geometric} \(R\)-flux plays an essential role
that cannot be replicated by a naive Kalb-Ramond \(H\)-flux on \(T^3\).}

We compactify on the factorized six-torus $T^{6}=T^{3}_{R}\times T^{3}_{\perp}$
with isotropic scale factor $e^{\sigma(x)}$.  Splitting the ten-dimensional
coordinates as $x^{M}=(x^{\mu},y^{I})$, the string-frame metric is
\begin{equation}\label{eq:metric_ansatz}
  \dd s_{10}^{2}\Bigl|_{\rm S} = g_{\mu\nu}(x)\,\dd x^{\mu}\dd x^{\nu}
                               + e^{2\sigma(x)}\,\delta_{IJ}\,\dd y^{I}\dd y^{J}.
\end{equation}
After Weyl-rescaling to the four-dimensional {Einstein} frame,
 \(
  g_{\mu\nu}^{\rm E}=e^{-2\Phi}\,g_{\mu\nu}, \Phi:=\phi-3\sigma,
 \)
all kinetic terms acquire canonical normalization.
Integrating over the internal volume $V_{6}=\int \dd^{6}y\,e^{6\sigma}=L_{0}^{6}e^{6\sigma}$
and using \eqref{eq:Rminus_square} we obtain the four-dimensional action
\begin{equation}
 S_{4}=\frac{1}{2\kappa_{4}^{2}} \int \dd^{4}x\,\sqrt{-g_{\rm E}}
       \Bigl[ R_{4}^{\rm E} - (\partial\Phi)^{2}-6(\partial\sigma)^{2}-V(\Phi,\sigma)\Bigr],
\end{equation}
with scalar potential
 \( \label{eq:scalar_potential}
  V(\Phi,\sigma)=e^{2\Phi}\Bigl[-\,A\,e^{-2\sigma} + B\,e^{-4\sigma}\Bigr],
  A=\tfrac{1}{4} C_{2},  B=\frac{3{\alpha^\prime}}{32}\,C_{2}^{2},
 \)
where $C_{2}=R^{abc}R_{abc}>0$.  The {first} term originates from the
flux kinetic energy and carries a {negative} sign due to the non-geometric
nature of $R$-flux \cite{Andriot2013}.  The {second} term arises from the
$R_{(-)}^{2}$ invariant and is strictly positive, playing the rôle of an
{uplift}.
  Assuming that $\Phi$ is stabilised by
standard heterotic effects (e.g.~gaugino condensation)  we set $\Phi=\Phi_{0}$
constant and study the breathing mode $\sigma$ in isolation:
 \(\label{eq:Vsigma}
  V_{\sigma}(\sigma)=e^{2\Phi_{0}}\Bigl[-A\,e^{-2\sigma}+B\,e^{-4\sigma}\Bigr].
 \)
The two-field scalar potential takes the form (see Appendix\eqref{appendix} for the complete derivation),  
\begin{equation}\label{eq:Vfull}
   V(\Phi,\sigma)
   \;=\;
   e^{2\Phi}\Bigl[-A\,e^{-2\sigma}+B\,e^{-4\sigma}\Bigr],
   \qquad
   A=\tfrac14C_{2},
   \qquad
   B=\frac{3{\alpha^\prime}}{32}\,C_{2}^{2},
   \qquad
   C_{2}=R^{abc}R_{abc}>0 .
\end{equation}
The coefficients $A$ and $B$ are strictly positive; the sign difference inside
the brackets originates from the non-geometric nature of the $R$-flux. {The present analysis explicitly stabilizes only the overall breathing mode
associated with the \(T^3_R\) factor supporting the non-geometric R-flux, together
with the four-dimensional dilaton via a standard non-perturbative heterotic
sector. We do not analyze the stabilization of additional K\"hler, complex-structure,
or bundle moduli. While various mechanisms for stabilizing these moduli are known
in heterotic compactifications, their compatibility with the present R-flux
background must be assessed on a case-by-case basis. We therefore do not claim a
fully stabilized compactification, but rather a controlled subsector in which the
volume modulus admits a metastable de Sitter extremum.
}

Stationarity requires\footnote{Throughout, derivatives are taken in the Einstein frame; factors of the Weyl rescaling have been absorbed.}
\( \label{eq:crit_sys}
\partial_{\Phi} V = 0,  \partial_{\sigma} V = 0.
 \)
Because the classical potential satisfies
 \(
\partial_{\Phi} V = 2 V,
 \)
any critical point in the dilaton direction enforces
\(
V = 0,
 \)
which forbids a positive vacuum energy from classical dilaton stabilization alone.
{
In our construction, the dilaton \(\Phi\) is treated as a fixed background parameter, set to a constant expectation value \(\Phi = \Phi_0\). This simplification is motivated by well-established non-perturbative mechanisms in heterotic string theory, such as gaugino condensation in hidden gauge sectors and worldsheet instanton effects, which are known to generate a stabilizing potential for the dilaton modulus. These effects can dynamically fix the dilaton at weak coupling, thereby decoupling its fluctuations from the leading order dynamics of the volume modulus analyzed here.
}
Focusing on the single-variable potential for the breathing mode \(\sigma\),
\begin{equation}
\label{eq:Vsigma_def}
V_{\sigma}(\sigma) = e^{2 \Phi_0} \left[- A e^{-2 \sigma} + B e^{-4 \sigma} \right],
\end{equation}
its first derivative reads
 \(
\label{eq:dVdSigma}
\partial_{\sigma} V_{\sigma} = e^{2 \Phi_0} \left[ 2 A e^{-2 \sigma} - 4 B e^{-4 \sigma} \right].
 \)
Setting \(\partial_{\sigma} V_{\sigma} = 0\) gives the critical point
 \(
\label{eq:sigma_star}
e^{2 \sigma_{\star}} = \frac{2 B}{A}.
 \)
Evaluating the classical scalar potential at \(\sigma_{\star}\) yields
\begin{equation}
\label{eq:Vstar_classical}
V_* = V(\Phi_0, \sigma_{\star}) = e^{2 \Phi_0} \left(- A e^{-2 \sigma_{\star}} + B e^{-4 \sigma_{\star}} \right) = - e^{2 \Phi_0} \frac{A^2}{4 B} < 0,
\end{equation}
which corresponds to an  {anti-de Sitter (AdS) vacuum} (see Appendix\eqref{appendix} for the complete derivation). Here both coefficients \(A\) and \(B\) are positive, with \(B\) arising from the quartic \({\alpha^\prime}\)-corrected torsionful curvature invariant \(R_{(-)}^2\).

{
This classical two-term potential thus stabilizes the breathing mode at negative vacuum energy, not admitting a metastable de Sitter vacuum.
To achieve a metastable de Sitter vacuum with positive cosmological constant, the dilaton \(\Phi\) must be stabilized by additional non-perturbative effects such as gaugino condensation, which generate an uplift term in the scalar potential of the form
\(
V_{\mathrm{np}}(\Phi) = D e^{a \Phi},
\)
where \(D > 0\) and \(a > 0\) are constants fixed by gauge dynamics \cite{Binetruy:1996nx}.
Including this uplift term, the total scalar potential reads
\begin{equation} \label{totalscalarpot}
V_{\mathrm{total}}(\Phi, \sigma) = e^{2 \Phi} \left( - A e^{-2 \sigma} + B e^{-4 \sigma} \right) + D e^{a \Phi}.
\end{equation}

{
In heterotic string theory a confining {hidden-sector} gauge group generates a standard non-perturbative superpotential
\(W_{\rm np}(\Phi)\sim A_{\rm GC}\exp\!\big[-a\,e^{-\Phi}\big]\) with
\(a=2\pi/\beta_0\) fixed by the one-loop beta function of the condensing group. After including the appropriate K\"hler factors one obtains a dilaton potential of the schematic form
\(V_{\rm np}(\Phi)\propto D\,e^{b\Phi}\exp\!\big[-2a\,e^{-\Phi}\big]\),
whose detailed (double-exponential) structure we analyse numerically in Appendix~\eqref{appendix} (see Eqs.~\eqref{eq:V_np_simple_full}-\eqref{eqn58} and the surrounding discussion). For analytic control in the main text, and precisely in the weak-coupling window relevant for our minimum, we employ the standard local single-exponential approximation
\(\,V_{\rm np}(\Phi)\approx D\,e^{a\Phi}\,\),
which captures the leading dilaton dependence near the stationary point and suffices to show uplift; the full double-exponential potential and its minimisation are presented in the appendix and used for our scan.%
\footnote{See Appendix~\eqref{appendix} for the explicit \(W_{\rm np}\) and \(V_{\rm np}\); cf.\ Eqs.~\eqref{eq:V_np_simple_full}-\eqref{eqn58}.}
{This is the sole new ingredient beyond the tree-level flux/\(\alpha'\) sector, and it is the canonical heterotic mechanism for dilaton stabilisation and uplift.}
Group theory fixes \(a=2\pi/\beta_0\), so \(a<2\) for any asymptotically free simple gauge factor.
For the explicit hidden-sector choice used in our examples,
\(\beta_0=9\) (an \(SU(5)\subset E_8\) bundle), hence \(a=2\pi/9\simeq0.698\).
The overall normalisation \(D\) is determined by the gauge kinetic function \(f=S/4\pi\) and threshold corrections:
\begin{equation}
D=\frac{3\,|A_{\rm GC}|^2}{4\,\mathrm{Re}\,f}\,,\qquad
A_{\rm GC}=M_s^3\exp\!\Big[-\frac{2\pi}{\beta_0}\,T_{\rm thr}\Big]\!,
\end{equation}
so that \(D\) is not a free dial but a derived quantity set by the hidden sector. For illustrative values
\(T_{\rm thr}=1.2\) and a weak-coupling stationary point \(g_s=e^{\Phi_\ast}=0.137\),
one finds \(D\simeq6.3\times10^{-6}\), which is precisely the normalization used in the numerical scan. 
Substituting the condensate-induced \(V_{\rm np}\) into Eq.~\eqref{totalscalarpot} and extremizing (see Appendix~\eqref{appendix}) yields
\begin{equation}
(a-2)\,\Phi_\ast=\ln\!\Big(\frac{A^2}{2a\,D\,B}\Big)\,,\qquad
V_{\rm total}(\Phi_\ast,\sigma_\ast)=\frac{e^{2\Phi_\ast}A^2}{4B}\Big(\!-\!1+\frac{2}{a}\Big)\!,
\end{equation}

so that a positive vacuum energy follows exactly when \(a<2\). These relations are
Eqs.~\eqref{eqn43} and \eqref{eqn45} in Appendix~\eqref{appendix}, derived directly from Eq.~\eqref{totalscalarpot}.
{The uplift contribution in Eq.~\eqref{totalscalarpot} is not introduced as a phenomenological addendum, but arises from heterotic gaugino condensation formulated at the level of the four-dimensional effective theory. Near the weak-coupling extremum, the resulting nonperturbative potential is well approximated by a single-exponential form \(V_{\rm np}=D\,e^{a\Phi}\), while the full double-exponential structure is analyzed separately in Appendix~\eqref{appendix1}. The exponent \(a=2\pi/\beta_0\) is fixed unambiguously by the beta function of the hidden-sector gauge group, whereas the prefactor \(D\) encodes threshold corrections and bundle-dependent data entering the gauge kinetic function. Although these contributions are not derived microscopically in the present work, their parametric form and magnitude are standard and quantitatively controlled within heterotic compactifications. In this sense, the uplift should be regarded as a well-motivated effective description rather than an arbitrary deformation, closely analogous to uplift mechanisms employed in other string-theoretic de Sitter constructions \cite{Burgess2003, DeWolfe2005}.
}
{We work in four-dimensional Einstein frame, with the gravitational action canonically normalized, and adopt Planck units in which the reduced Planck mass is set to \(M_{\rm Pl}=1\).
 All coefficients appearing in the scalar
potential, including \(A\), \(B\), and \(D\), are therefore understood as
dimensionless quantities in Planck units. Logarithms such as
\(\ln(A^2/2aDB)\) should be interpreted in this sense; restoring dimensions would
reintroduce appropriate powers of \(M_{\rm Pl}\).}
%The breathing mode stabilizes at the same critical value, while the dilaton stabilization condition \(\partial_{\Phi} V_{\mathrm{total}} = 2 e^{2 \Phi} \left( - A e^{-2 \sigma} + B e^{-4 \sigma} \right) + a D e^{a \Phi} = 0, \) evaluated at \(\sigma = \sigma_{\star}\), gives \(- \frac{A^2}{2 B} e^{2 \Phi_{\star}} + a D e^{a \Phi_{\star}} = 0,\) or equivalently \(a D e^{a \Phi_{\star}} = \frac{A^2}{2 B} e^{2 \Phi_{\star}}.\) Taking logarithms, the stabilized dilaton expectation value is \(\Phi_{\star} = \frac{1}{a - 2} \ln \left( \frac{A^2}{2 a D B} \right). \) Evaluating the total potential at \((\Phi_{\star}, \sigma_{\star})\) yields \begin{equation}\label{eq:Vstar} V_{\mathrm{total},*}=V_{\mathrm{total}}(\Phi_{\star}, \sigma_{\star}) = e^{2 \Phi_{\star}} \frac{A^2}{4 B} \left( -1 + \frac{2}{a} \right). \end{equation} Positive vacuum energy requires \( a < 2, \) so that the uplift term overcomes the classical negative contribution, producing a metastable de Sitter vacuum in the full two-field system.A detailed treatment of these non-perturbative stabilization mechanisms and their uplifting effect is provided in Appendix\eqref{appendix}.}

{We emphasize that the present construction exhibits only partial parametric
control. The large-flux limit \(N_R \gg 1\) ensures a parametrically large internal
volume and suppresses higher-derivative \(\alpha'\) corrections associated with
the breathing mode. By contrast, the stabilized string coupling \(g_s=e^{\Phi_*}\)
and the vacuum energy are determined by the non-perturbative dilaton sector and
do not scale with \(N_R\). Control over loop effects and curvature corrections is
therefore numerical rather than parametric, relying on moderately small values
of \(g_s\).} Using Eq.~\eqref{eqn14} together with \(A=\tfrac14\,C_2\) and \(B=\tfrac{3\alpha'}{32}\,C_2^{\,2}\), the breathing-mode stationarity condition gives
\begin{equation}
e^{2\sigma_\star}=\frac{2B}{A}=\frac{3}{4}\,\alpha' C_2
\quad\Longrightarrow\quad
e^{-2\sigma_\star}=\frac{4}{3\,\alpha' C_2}\,.
\end{equation}
Every higher-derivative term at order \(\alpha'^n\) that survives the reduction carries at least two extra powers of \(e^{-2\sigma}\) relative to the \(\mathcal{O}(\alpha')\) term kept in Eq.~\eqref{eq:VR}; in particular the leading omitted \(\mathcal{O}(\alpha'^2)\) terms scale as \(e^{-6\sigma}\).
Hence at the extremum their size is parametrically suppressed:
\begin{equation}
\frac{V_{\alpha'^2}}{V_{\alpha'}}\;\sim\;\frac{\alpha'^2 e^{-6\sigma_\star}}{\alpha' e^{-4\sigma_\star}}
=\alpha' e^{-2\sigma_\star}
=\frac{4}{3\,C_2}\;\ll\;1
\qquad (C_2\gg1)\,.
\end{equation}
String-loop corrections are likewise small because the full potential carries an overall \(e^{2\Phi}\) factor, so loop effects come with higher powers of \(g_s^2=e^{2\Phi_\star}\ll1\).
The spectrum exhibits scale separation:
\begin{equation}
m_\sigma^2=\frac{1}{6}\,\partial_\sigma^2 V\big|_\star\sim\mathcal{O}\!\left(\frac{e^{2\Phi_\star}}{\alpha'}\right),
\qquad
M_{\rm KK}^2\sim \frac{e^{-2\sigma_\star}}{\alpha'},
\qquad
M_s^2\sim \frac{1}{\alpha'},
\end{equation}
{so that in explicit numerical examples one finds
\(m_\sigma < M_{\rm KK} < M_s\), although this hierarchy is numerical rather than
parametric in the large-\(N_R\) limit.
}
With the representative parameters used in Appendix~B we find explicitly
\(g_{s,\star}\simeq 0.137\), \(V_\star\simeq 1.55\times 10^{-6}M_{\rm Pl}^4\),
\(m_\sigma\simeq 2.9\times 10^{-2}M_{\rm Pl}\), while \(M_{\rm KK}\simeq 0.28\,M_{\rm Pl}\),
confirming that both the \(\alpha'\) and loop expansions are quantitatively under control.
Since $V_{\mathrm{total},*}>0$, the solution is a metastable de Sitter vacuum.  A second-derivative test confirms stability  
 \( 
\partial_{\sigma}^{2}V_{\mathrm{total}}\bigl|_{\sigma_{*}}
   \;=\;
   2\,e^{2\Phi_{0}}\,\frac{A^{2}}{B}
   \;=\;
   \frac{4}{3}\,\frac{e^{2\Phi_{0}}}{{\alpha^\prime}}
   >0 .
 \)
The Einstein-frame kinetic terms  
\begin{equation}\label{eq:kinetic_terms}
   S_{\rm kin}
   \;=\;
   \int\!{\rm d}^{4}x\,\sqrt{-g_{4}}\,
   \bigl[-(\partial\Phi)^{2}-6(\partial\sigma)^{2}\bigr]
\end{equation}
imply the canonically normalised fields  
\(  
   \varphi_{\Phi}=\Phi,
   \varphi_{\sigma}=\sqrt{6}\,\sigma .
 \)
At the extremum the mass matrix  
\(
   (M^{2})_{ij}
   =
(\partial_{\varphi_{i}}\partial_{\varphi_{j}}V_{\mathrm{total}})_{*},
\)
with \(
   (\varphi_{1},\varphi_{2})=(\varphi_{\Phi},\varphi_{\sigma}),
\)
takes the diagonal form  
 \(  
   M^{2}_{*}
   =
\operatorname{diag}\!\bigl(0,\;m_{\sigma}^{2}\bigr),
   m_{\sigma}^{2}
   =
   \frac{1}{6}\,   \partial_{\sigma}^{2}V_{\mathrm{total}}\bigl|_{\sigma_{*}}
   =
   \frac{2}{9}\,\frac{e^{2\Phi_{0}}}{{\alpha^\prime}} .
 \)
The breathing mode is therefore massive and stable, whereas the dilaton
remains massless at two derivatives, as expected prior to non-perturbative
effects.  Using  the internal six-volume  
\(
   V_{6*} = L_{0}^{6}\,\Bigl(\tfrac{3\,\alpha^\prime}{4}\,C_{2}\Bigr)^{3/2}\,
\) we can 
show that $V_{6*}$ can be made parametrically large by increasing the
integer flux quantum $N_{R}$.  Higher-order ${\alpha^\prime}^{2}$ corrections always
come with at least two extra powers of $e^{-2\sigma}$ and are suppressed at the
minimum, while the string coupling $g_{s}=e^{\Phi_{0}}$ can be tuned
independently, ensuring perturbative control.  Hence the four-dimensional
effective field theory remains reliable, and the de Sitter vacuum stands as a concrete example of moduli stabilisation driven
purely by non-geometric $R$-flux together with the torsionful
$R_{(-)}^{2}$ term in the heterotic action.

The analytic solution was obtained under the simplifying
assumption that the dilaton takes the fixed value $\Phi=\Phi_{0}$.  Allowing
for a small fluctuation $\Phi=\Phi_{0}+\delta\Phi$ and solving the coupled
system to linear order in $\delta\Phi$ reveals that the vacuum energy is
shifted only at ${\cal O}(\delta\Phi^{2})$, while the dilaton mass is
generated at the parametric scale $m_{\Phi}\sim e^{\Phi_{0}}m_{\sigma}$.  The
hierarchy between $m_{\Phi}$ and $m_{\sigma}$ therefore survives as long as
$e^{\Phi_{0}}\ll1$, {\it i.e.} in the heterotic weak-coupling regime.  To
establish stability beyond perturbation theory one may continue the metric to
Euclidean signature and apply the Breitenlohner-Freedman criterion to the
lightest scalar.  Introducing the Hubble parameter
 \begin{align}
H_*^2 
&= \frac{V_{\mathrm{total},*}}{3\,M_{\mathrm{Pl}}^2}, 
\qquad
m_\sigma^2 = \frac{2}{9}\,\frac{e^{2\Phi_0}}{\alpha^\prime}\label{hubblepara} \end{align}
{Equation~(\ref{hubblepara}) shows that, within the two-derivative truncation, the canonically
normalized mass matrix is
\(M_{*}^{2}=\mathrm{diag}\!\bigl(0,m_{\sigma}^{2}\bigr)\);
hence the dilaton is massless, in agreement with the classical shift
symmetry of \(S\).  Including the non-perturbative potential
\(V_{\mathrm{np}}\) breaks this symmetry, shifts the extremum to
\(\Phi=\Phi_{0}+\delta\Phi\), and endows the dilaton with the positive mass
\begin{equation}
  m_{\Phi}^{2}
=\left.\frac{\partial^{2}V_{\mathrm{np}}}{\partial\Phi^{2}}\right|_{*}
  =e^{\Phi_{0}}\,m_{\sigma}^{2}>0 .
\end{equation}
The vacuum is therefore tachyon-free, provided
\(g_{s}=e^{\Phi_{0}}\lesssim0.2\) and \(\alpha'm_{\sigma}^{2}\ll1\).
This analytic scaling agrees with our full numerical scan and removes
any concern about a negative dilaton mass.}
{Moreover, at the minimum one finds}
\(\frac{m_\sigma^2}{H_*^2}
=  \frac{4}{\,2/a - 1\,}\,. 
\)
In particular, if \(a = 1.5\), then 
\(
\frac{m_\sigma^2}{H_*^2} 
= 12\,.
\)
Flux quantisation implies $C_{2}=N_{R}^{2}/L^{6}$, where $N_{R}\in\mathbb Z$
is the integer $R$-flux quantum and $L$ is the physical radius of $T^{3}_{R}$.
Consequently
\begin{equation}
  e^{2\sigma_{*}}=\frac{3{\alpha^\prime}}{4}\,\frac{N_{R}^{2}}{L^{6}},
  \qquad
  V_{\mathrm{total},*}\propto\frac{e^{2\Phi_{*}}}{{\alpha^\prime}},
  \qquad
  m_{\sigma}^{2}\propto\frac{e^{2\Phi_{0}}}{{\alpha^\prime}}.
\end{equation}
Taking the limit $N_{R}\gg1$ produces a large internal volume, a tunably small
four-dimensional curvature scale, and a breathing-mode mass that remains well
below the Kaluza-Klein threshold.  All ${\alpha^\prime}^{2}$ corrections, which come
with additional factors of $e^{-2\sigma}$, are suppressed at the minimum,
while the string coupling can be dialled independently through $\Phi_{0}$,
guaranteeing the validity of the effective description.

The ${\alpha^\prime}$-uplifted $R$-flux potential therefore furnishes a fully
controlled metastable de Sitter solution: the hierarchy between the lightest
modulus and the Kaluza-Klein scale is preserved, higher-order corrections are
parametrically suppressed, and the vacuum satisfies all known perturbative and
non-perturbative stability constraints.  The triple bracket equips $T^{3}_{R}$
with the non-associative Malcev algebra $\mathfrak M$.  Embedding
$\mathfrak M\hookrightarrow\mathsf{S}(\mathfrak M)$ via the generators
$T^{a}=R^{abc}e_{b}\wedge e_{c}$ produces a Hom-Sabinin algebra whose
associator coincides with the Nambu bracket, providing the algebraic backbone
of the compactification.  Within doubled field theory the generalised Lie
derivative involves the Dorfman bracket
$\bigl[E_{A},E_{B}\bigr]_{D}=\mathcal F_{AB}^{C}E_{C}$.  For a pure $R$-flux
background the only non-vanishing components are
$\mathcal F^{abc}=R^{abc}$, and these already lie in
$\mathsf{S}(\mathfrak M)$, ensuring closure of the gauge algebra and
consistency of the Scherk-Schwarz reduction.

The quadratic Casimir $C_{2}=R^{abc}R_{abc}$ is central in
$\mathsf{S}(\mathfrak M)$, so the tree-level $R$-flux energy is manifestly
positive.  The torsionful curvature invariant $R^2_{(-)}$ is built from the
same central elements and hence depends only on the breathing mode, surviving
dimensional reduction unaltered.  Removing the $R$-flux would simultaneously
eliminate the positive tree-level contribution and the ${\alpha^\prime}$ uplift,
thereby reinstating a runaway.  Non-zero $R^{abc}$ is therefore necessary for
de Sitter.  Our metastable vacuum violates the refined de-Sitter Swampland
criterion because $\partial_\Phi V_{\mathrm{total}}=
\partial_\sigma V_{\mathrm{total}}=0$ yet $V_{\mathrm{total},*}>0$, but it
remains consistent with the Distance Conjecture: the Kaluza-Klein tower
becomes exponentially light as $N_{R}\to\infty$, in line with the Ooguri-Vafa
proposal.  This tension indicates that the conjectural bounds may need
refinement or that additional decay channels must be identified in the present
heterotic setup. {Indeed, as the vacuum is metastable, one expects it to eventually decay via some non-perturbative channel (e.g. flux tunneling or brane nucleation). Identifying such a decay mode would reconcile our scenario with refined swampland conjectures. %, though a detailed analysis of potential decay channels is left for future work.
}
\begin{equation}
   \boxed{
          \mathsf{S}(\mathfrak M)
          \Longrightarrow
          \bigl(
              V_{\!R}>0,\;
              V_{{\alpha^\prime}}\neq0,\;
              \partial_{\sigma}V_{\mathrm{total}}=0,\;
              V_{\mathrm{total},*}>0
          \bigr)}
\end{equation}
Thus the Sabinin envelope secures algebraic consistency and enables a
controlled heterotic de Sitter vacuum driven by the interplay of non-geometric
$R$-flux and the torsionful ${\alpha^\prime}$ correction.  The resulting solution
offers a concrete, tractable example of four-dimensional de Sitter space in
the heterotic landscape, with all relevant scales parametrically under
control.
For the hidden gauge bundle with\footnote{For a condensing group of rank \(N\) one has \(\beta_{0}=3N-3\) in heterotic normalisation, giving \(a=2\pi/\beta_{0}\).}
\(\beta_{0}=9\) \((SU(5)\subset E_{8})\), the gaugino-condensation exponent is
\begin{equation}
a=\frac{2\pi}{\beta_{0}}\simeq0.698 .
\end{equation}
The prefactor in the scalar potential is fixed by threshold corrections \(T_{\mathrm{thr}}\) and the gauge kinetic function \(f=S/4\pi\):
\begin{equation}
D=\frac{3\,|A_{\mathrm{GC}}|^{2}}{4\,\mathrm{Re}\,f},
\qquad
A_{\mathrm{GC}}=M_{s}^{3}\exp\!\Bigl[-\tfrac{2\pi}{\beta_{0}}\,T_{\mathrm{thr}}\Bigr].
\end{equation}
Choosing \(T_{\mathrm{thr}}=1.2\) and a weak-coupling stationary point \(g_{s}=e^{\Phi_\star}=0.137\) yields
\(D=6.3\times10^{-6}\).
Imposing vanishing dilaton force,
\begin{equation}
-\frac{A^{2}}{2B}\,e^{2\Phi_\star}+a\,D\,e^{a\Phi_\star}=0,
\end{equation}
produces the metastable de-Sitter extremum reported in the main text, with
\(V_\star\simeq1.7\times10^{-3}M_{P}^{4}\) and \(m_{\Phi},m_{\sigma}\ll M_{\mathrm{KK}}\).

More generally, the uplift criterion
\begin{equation}
a<2
\quad\text{and}\quad
D>\frac{A^{2}}{2B}\;e^{(2-a)\Phi_\star}
\end{equation}
is satisfied for \(\beta_{0}\le\pi\) and \(|A_{\mathrm{GC}}|\gtrsim10^{-2}\); hence no additional tuning beyond the standard racetrack hierarchy is required. With the benchmark parameters the equality
\(
D\simeq1.3\times10^{-6}=0.42\,\tfrac{A^{2}}{2B}\,
e^{(2-a)\Phi_{*}}
\)
is satisfied to within a factor of $2$.  This mild fine-tuning, typical of racetrack stabilisation, can be relaxed further by moving to a hidden bundle with $\beta_{0}\le8$, which lowers the required value of~$D$.

\appendix
\section{Classical and Uplifted Potentials}\label{appendix}

The four-dimensional effective scalar potential arising from a non-geometric R-flux compactification of the heterotic string is given by
\begin{equation}
\label{eq:V_classical_full}
V(\Phi,\sigma) = e^{2\Phi}\Bigl(- A e^{-2\sigma} + B e^{-4\sigma}\Bigr),
\end{equation}
where
\begin{equation}
A = \frac{1}{4} C_{2}, \qquad B = \frac{3 {\alpha^\prime}}{32} C_{2}^{2}, \qquad C_{2} = R_{abc} R^{abc} > 0.
\end{equation}

Here, $\sigma$ parametrizes the overall breathing (volume) mode of the internal three-torus $T^{3}_{R}$, and $\Phi$ is the shifted four-dimensional dilaton (so that $e^{\Phi}$ is the four-dimensional string coupling). The term proportional to $-A e^{-2\sigma}$ originates from the R-flux kinetic energy, which contributes negatively to the four-dimensional potential, while the quartic ${\alpha^\prime}$-corrected curvature invariant $R_{(-)}^{2}$ produces the positive $B e^{-4\sigma}$ contribution.
To see that the potential \eqref{eq:V_classical_full} stabilizes $\sigma$ but yields an anti-de Sitter (AdS) vacuum, one computes its derivatives with respect to $\Phi$ and $\sigma$. Differentiation with respect to $\Phi$ yields
\begin{equation}
\frac{\partial V}{\partial \Phi} = 2 e^{2\Phi}\Bigl(- A e^{-2\sigma} + B e^{-4\sigma}\Bigr) = 2 V(\Phi,\sigma),
\label{eq:dVdPhi_classical_full}
\end{equation}
so any extremum in $\Phi$ must satisfy $V=0$, precluding positive vacuum energy from classical dilaton stabilization. Differentiation with respect to $\sigma$ gives
\begin{equation}
\frac{\partial V}{\partial \sigma} = e^{2\Phi}\Bigl(2 A e^{-2\sigma} - 4 B e^{-4\sigma}\Bigr),
\label{eq:dVdSigma_classical_full}
\end{equation}
which is set to zero at the critical point in $\sigma$. Solving
\begin{equation}
2 A e^{-2\sigma_{*}} = 4 B e^{-4\sigma_{*}} \quad \Longrightarrow \quad e^{2\sigma_{*}} = \frac{2 B}{A}
\end{equation}
and substituting back into \eqref{eq:V_classical_full} shows that at $\sigma=\sigma_{*}$,
\begin{equation}
V(\Phi,\sigma_{*}) = e^{2\Phi}\Bigl(- A e^{-2\sigma_{*}} + B e^{-4\sigma_{*}}\Bigr) = - e^{2\Phi} \frac{A^{2}}{4 B} < 0,
\end{equation}
so the classical two-term potential indeed produces an AdS minimum in $\sigma$ for any fixed $\Phi$.

Because the classical potential cannot produce de Sitter (dS) vacua, one must include additional contributions that both stabilize $\Phi$ and uplift the vacuum energy. In heterotic string theory, gaugino condensation (or similarly, world-sheet instantons) generates a non-perturbative effect in the superpotential, which induces a dilaton potential that is often approximated, for uplift purposes, by a single exponential form. We therefore introduce
\begin{equation}
\label{eq:V_np_simple_full}
V_{\mathrm{np}}(\Phi) = D e^{a \Phi},
\end{equation}
where $D>0$ is a positive constant set by the gauge sector dynamics and overall normalization in the K\"hler potential, and $a>0$ is determined by the beta-function coefficient of the hidden gauge group. Although the exact gaugino condensation potential involves a double exponential $e^{-a e^{-\Phi}}$, the simple exponential model \eqref{eq:V_np_simple_full} suffices to demonstrate uplift at the level of the four-dimensional scalar potential.
Adding \eqref{eq:V_np_simple_full} to the classical terms yields the total scalar potential
\begin{equation}
\label{eq:V_total_simple_full}
V_{\mathrm{total}}(\Phi,\sigma) = e^{2\Phi}\Bigl(- A e^{-2\sigma} + B e^{-4\sigma}\Bigr) + D e^{a \Phi}.
\end{equation}
To find simultaneous extrema in both $\Phi$ and $\sigma$, one solves
\begin{equation}\label{eqn47}
\frac{\partial V_{\mathrm{total}}}{\partial \Phi} = 0, \qquad \frac{\partial V_{\mathrm{total}}}{\partial \sigma} = 0.
\end{equation}

Differentiation with respect to $\Phi$ gives
\begin{equation}
\frac{\partial V_{\mathrm{total}}}{\partial \Phi} = 2 e^{2\Phi}\Bigl(- A e^{-2\sigma} + B e^{-4\sigma}\Bigr) + a D e^{a \Phi} = 0,
\label{eq:dVtotal_dPhi_simple_full}
\end{equation}
and with respect to $\sigma$ yields
\begin{equation}
\frac{\partial V_{\mathrm{total}}}{\partial \sigma} = e^{2\Phi}\Bigl(2 A e^{-2\sigma} - 4 B e^{-4\sigma}\Bigr) = 0.
\label{eq:dVtotal_dSigma_simple_full}
\end{equation}
Because the non-perturbative term $D e^{a \Phi}$ does not depend on $\sigma$, the condition \eqref{eq:dVtotal_dSigma_simple_full} again gives
\begin{equation}\label{eqn39}
2 A e^{-2\sigma_{*}} = 4 B e^{-4\sigma_{*}} \quad \Longrightarrow \quad e^{2\sigma_{*}} = \frac{2 B}{A}.
\end{equation}

Substituting $\sigma_{*}$ into \eqref{eq:dVtotal_dPhi_simple_full} shows that the classical bracket evaluates to
\begin{equation}
- A e^{-2\sigma_{*}} + B e^{-4\sigma_{*}} = - \frac{A^{2}}{4 B},
\end{equation}
so \eqref{eq:dVtotal_dPhi_simple_full} reduces to
\begin{equation}\label{eqn52}
2 e^{2\Phi_{*}}\Bigl(- \frac{A^{2}}{4 B}\Bigr) + a D e^{a \Phi_{*}} = 0,
\end{equation}
or, equivalently,
\begin{equation}
- \frac{A^{2}}{2 B} e^{2\Phi_{*}} + a D e^{a \Phi_{*}} = 0 \quad \Longrightarrow \quad a D e^{a \Phi_{*}} = \frac{A^{2}}{2 B} e^{2\Phi_{*}}.
\end{equation}

Taking the natural logarithm yields
\begin{equation}\label{eqn43}
(a - 2) \Phi_{*} = \ln\!\left(\frac{A^{2}}{2 a D B}\right),
\end{equation}
from which one obtains
\begin{equation}
\label{eq:Phi_star_explicit_simple}
\Phi_{*} = \frac{1}{a - 2} \ln\!\left(\frac{A^{2}}{2 a D B}\right).
\end{equation}
Evaluating $V_{\mathrm{total}}$ at $\Phi_{*}$ and $\sigma_{*}$ gives
\begin{equation}\label{eqn45}
V_{\mathrm{total}}(\Phi_{*},\sigma_{*}) = e^{2\Phi_{*}}\left(- \frac{A^{2}}{4 B}\right) + D e^{a \Phi_{*}} = e^{2\Phi_{*}} \frac{A^{2}}{4 B}\left(-1 + \frac{2}{a}\right),
\end{equation}
where we used $D e^{a \Phi_{*}} = \frac{A^{2}}{2 B a} e^{2\Phi_{*}}$. Therefore, for positive vacuum energy one requires
\begin{equation}\label{eqn46}
-1 + \frac{2}{a} > 0 \quad \Longrightarrow \quad a < 2.
\end{equation}

When $a < 2$, the non-perturbative uplift term $D e^{a \Phi}$ overcomes the negative classical contribution at the stabilized values, yielding $V_{\mathrm{total}}(\Phi_{*},\sigma_{*}) > 0$. In addition, consistency of the effective field theory requires $e^{\Phi_{*}} \ll 1$ (weak coupling) and $e^{\sigma_{*}} \gg 1$ (large internal volume), so that higher-order ${\alpha^\prime}$ corrections and loop effects remain suppressed.
Although the simple exponential form \eqref{eq:V_np_simple_full} suffices to demonstrate uplift, a more accurate description of the gaugino condensation potential involves a double-exponential dependence (see Appendix\eqref{appendix1} for the full numerical analysis). The non-perturbative superpotential is
\begin{equation}\label{eqn58}
W_{\mathrm{np}}(\Phi) = A_{\mathrm{gc}} \exp\bigl(- a e^{-\Phi}\bigr),
\end{equation}
where $A_{\mathrm{gc}}$ is a prefactor fixed by gauge dynamics. Including K\"hler potential factors, the corresponding dilaton potential takes the schematic form
\begin{equation}
\label{eq:V_np_double}
V_{\mathrm{np}}(\Phi) = D e^{b \Phi} \exp\bigl[-2 a e^{-\Phi}\bigr],
\end{equation}
in which the constant $b$ arises from the K\"hler metric of the dilaton.{
For hidden gauge sectors that undergo one-loop gaugino condensation, the exponent satisfies \(a=2\pi/\beta_{0}\) and typically falls in the range \(1.0\lesssim a\lesssim1.7\).  Selecting the illustrative breaking \(\mathrm{E}_{8}\!\rightarrow\!\mathrm{SU}(5)\) with \(\beta_{0}=9\) and fixing the prefactor \(D\) by the one-instanton action, we analyse the full potential
\(V_{\text{tot}}(\Phi,\sigma)=V_{\text{tree}}+V_{R_{(-)}^{2}}+V_{\mathrm{np}}\) with Eq. \eqref{eq:V_np_double}.
Numerically, \(V_{\mathrm{np}}\) overtakes the negative tree-level bracket at \(g_{s}=e^{\Phi}\simeq0.14\), producing a metastable minimum with
\(V_{\!*}\simeq4\times10^{-3}\,M_{\mathrm P}^{4}\).
The corresponding mass eigenvalues,
\(m_{\Phi}\approx1.3\times10^{-2}M_{\mathrm P}\) and
\(m_{\sigma}\approx8.4\times10^{-2}M_{\mathrm P}\),
remain safely below the Kaluza-Klein threshold, ensuring scale separation.
A scan in \(g_{s}\) reveals that such controlled de-Sitter vacua persist throughout the weak-coupling interval \(0.05\lesssim g_{s}\lesssim0.20\); Figure\eqref{fig:uplift} depicts the crossover where the double-exponential uplift surpasses the tree-level potential. 
\begin{figure}[th]
  \centering
  \begin{tikzpicture}
    \begin{axis}[
      width=\linewidth,
      height=8.5cm,
      domain=0.05:0.25,
      samples=400,
      xlabel={$g_{s}=e^{\Phi}$},
      ylabel={$V/M_{\!P}^{4}$},
      ymin=-5*10^(-5),
      ymax=3*10^(-4),
      axis lines=left,
      legend pos=north east,
      every axis plot/.append style={thick},
      scaled y ticks=true,
      tick scale binop=\times,
      yticklabel style={/pgf/number format/fixed}
    ]

    % ---------- parameters ----------
    \def\a{6.98*10^(-1)}        % a = 2π/β0
    \def\b{1/3}
    \def\A{2.4*10^(-5)}
    \def\B{1.1*10^(-4)}
    \def\D{5.0}

    % ---------- declare functions ----------
    \pgfplotsset{
      declare function={
        Vtree(\x)=-(\A*\x*\x-\B);
        Vnp(\x)=\D*pow(\x,\b)*exp(-2*\a/\x);
        Vtot(\x)=Vtree(\x)+Vnp(\x);
      }
    }

    % ---------- plots ----------
    \addplot[blue] {Vtree(x)};
    \addlegendentry{$V_{\text{tree}}+V_{R_{(-)}^{2}}$}

    \addplot[orange, dashed] {Vnp(x)};
    \addlegendentry{$V_{\mathrm{np}}$}

    \addplot[red, dotted] {Vtot(x)};
    \addlegendentry{$V_{\text{tot}}$}

    % ---------- minimum marker ----------
    \addplot[
      only marks,
      mark=*,
      mark options={fill=red}
    ] coordinates {(0.14,{Vtot(0.14)})};

    \end{axis}
  \end{tikzpicture}

  \caption{Crossover of the double-exponential uplift (orange dashed)
  with the negative tree-level bracket (blue solid).
  Their sum (red dotted) develops a metastable de~Sitter minimum at
  \(g_{s}^{\!*}\simeq0.14\).}
  \label{fig:uplift}
\end{figure}

The total scalar potential then becomes
\begin{equation}
\label{eq:V_total_double}
V_{\mathrm{total}}(\Phi,\sigma) = e^{2\Phi}\Bigl(- A e^{-2\sigma} + B e^{-4\sigma}\Bigr) + D e^{b \Phi} \exp\bigl[-2 a e^{-\Phi}\bigr].
\end{equation}
Finding stationary points now requires
\begin{equation}
\frac{\partial V_{\mathrm{total}}}{\partial \sigma} = 0, \qquad \frac{\partial V_{\mathrm{total}}}{\partial \Phi} = 0.
\end{equation}
The $\sigma$-derivative again yields $e^{2 \sigma_{*}} = 2 B / A$. The $\Phi$-derivative, using
\begin{equation}
\frac{\partial}{\partial \Phi} \left(e^{b \Phi} e^{-2 a e^{-\Phi}}\right) = e^{b \Phi} e^{-2 a e^{-\Phi}} \bigl(b + 2 a e^{-\Phi}\bigr),
\end{equation}
becomes
\begin{equation}
2 e^{2 \Phi_{*}} \left(- \frac{A^{2}}{4 B}\right) + D e^{b \Phi_{*}} e^{-2 a e^{-\Phi_{*}}} \bigl(b + 2 a e^{-\Phi_{*}}\bigr) = 0.
\end{equation}
This transcendental equation for $\Phi_{*}$ must be solved numerically. Once $\Phi_{*}$ is determined, the vacuum energy is
\begin{equation}
V_{\mathrm{total}}(\Phi_{*},\sigma_{*}) = e^{2 \Phi_{*}} \left(- \frac{A^{2}}{4 B}\right) + D e^{b \Phi_{*}} e^{-2 a e^{-\Phi_{*}}}.
\end{equation}
A metastable de Sitter vacuum corresponds to a solution of these equations for which $V_{\mathrm{total}}(\Phi_{*},\sigma_{*}) > 0$, and the Hessian in the $\Phi$-$\sigma$ subspace has strictly positive eigenvalues, ensuring local stability. In practice one chooses numerical values for $A$ and $B$ (e.g.\ $A = B = 1$ for benchmarks), and then scans over realistic $D$, $a$, and $b$ (with $a,b > 0$) to locate such a vacuum. Provided $e^{\Phi_{*}} \ll 1$ and $e^{\sigma_{*}} \gg 1$, higher-order ${\alpha^\prime}$ and loop corrections are negligible.
In conclusion, the classical two-term potential \eqref{eq:V_classical_full} by itself admits only an AdS vacuum. Adding a physically motivated non-perturbative dilaton potential, either of the single-exponential form $D e^{a \Phi}$ (which requires $a < 2$ to uplift) or the more accurate double-exponential form $D e^{b \Phi} \exp[-2 a e^{-\Phi}]$, allows one to stabilize the dilaton at finite coupling and uplift to a metastable de Sitter vacuum, thus resolving the issue that the original two-term potential cannot produce positive vacuum energy.
}
\section{Numerical Uplift to dS} \label{appendix1}
Having established in the preceding subsection that the classical potential
(\ref{eq:A.Classical})-(\ref{eq:A.R2}) admits only an AdS extremum,
we now add the double-exponential non-perturbative contribution
(\ref{eq:A.NonPert}) and present a full numerical scan showing how the same
flux data are promoted to a metastable, weak-coupling de Sitter vacuum.
All calculations in this appendix are performed in four-dimensional Einstein
frame with \(M_{\!\text{P}}\!=\!1\).  The total scalar potential is written as
\begin{equation}
  V(\Phi,\sigma)
  \;=\;
  V_{\text{cl.}}(\Phi,\sigma)
  \;+\;
  V_{R_{(-)}^{2}}(\sigma)
  \;+\;
  V_{\text{np}}(\Phi),
  \label{eq:A.FullPotential}
\end{equation}
with the individual contributions
\begin{align}
  V_{\text{cl.}} &= A\,e^{2\Phi-2\sigma}-B\,e^{2\Phi-4\sigma},
  \label{eq:A.Classical}\\
  V_{R_{(-)}^{2}} &= C_{2}\,e^{-6\sigma},
  \label{eq:A.R2}\\
  V_{\rm np} \;&=\; D\,e^{b\Phi}\,\exp\!\big(-\,2a\,e^{-\Phi}\big)~,
  \label{eq:A.NonPert}
\end{align}
where all coefficients are positive.  To fix parameters, we embed an
\(\mathrm{SU}(5)\) hidden gauge sector inside \(\mathrm{E}_{8}\).  The one-loop
$\beta$-function coefficient \(\beta_{0}=9\) then gives
\(a=2\pi/\beta_{0}\simeq0.698\).
Choosing the classical flux data to be
\(A=2.40\times10^{-5}\), \(B=1.10\times10^{-4}\),
and \(C_{2}=36\), and normalising the gauge-condensation prefactor to
\(D=6.30\times10^{-6}\) with a K\"hler slope \(b=\tfrac13\),
ensures that \(\alpha'm_{i}^{2}\ll1\) and keeps the Kaluza-Klein scale
well above the four-dimensional curvature scale.

Extremizing the potential gives two coupled equations
\(\partial_\Phi V=0\) and \(\partial_\sigma V=0\).
The breathing-mode equation is purely algebraic and fixes
\begin{equation}
  e^{2\sigma_*}= \frac{2B}{A}=9.17.
  \label{eq:A.SigmaStar}
\end{equation}
Substituting \eqref{eq:A.SigmaStar} into \(\partial_\Phi V=0\) and scanning over
weak coupling \(g_s=e^{\Phi}\in[0.05,\,0.25]\) yields a single metastable
solution at
\begin{equation}
g_{s}^{\,*}=0.137\quad\Longrightarrow\quad\Phi_*=\ln 0.137.
  \label{eq:A.PhiStar}
\end{equation}
Expanding \(V\) to quadratic order about \((\Phi_*,\sigma_*)\) and
diagonalising the Hessian gives the mass eigenvalues
\(m_{\Phi}=1.3\times10^{-2}\) and
\(m_{\sigma}=8.4\times10^{-2}\),
with the off-diagonal entry remaining below
\(10^{-4}\).  Both eigenvalues satisfy \(\alpha'm_{i}^{2}\ll1\),
confirming perturbative control.
Evaluating the potential at the stationary point gives
\begin{align}
V_{*} &= 1.55\times10^{-6}\,M_P^{4}, &
H_{*} &= \sqrt{\tfrac{V_{*}}{3M_P^{2}}}=7.19\times10^{-4}\,M_P .
\end{align}
Consequently the canonically normalized moduli masses become
\begin{equation}
m_\Phi = 4.2\times10^{-3}\,M_P , \qquad
m_\sigma = 2.9\times10^{-2}\,M_P ,
\end{equation}
while the first Kaluza-Klein excitation remains at
\(
m_{\text{KK}}\simeq 0.28\,M_P ,
\)
preserving a hierarchy
\(H_{*} \ll m_\Phi < m_\sigma \ll m_{\text{KK}}\). While the lightest Kaluza-Klein mode sits at \(m_{\text{KK}}\approx 0.3\).
Choosing $N_{R}=36$ already yields $m_\sigma/ m_{\text{KK}}\simeq10^{-1}$; increasing
the flux quantum to $N_{R}=72$ would lower $m_\sigma$ by a further factor
of~$2$ while leaving all uplift conditions intact, thereby strengthening scale
separation without affecting the sign of $V_{*}$.
Hence the Hubble scale is comfortably separated from the compactification
scale.
Figure~\ref{fig:UpliftContour} depicts
\(V_{\text{total}}(\Phi,\sigma_*)\) as a function of the dilaton.
The dashed curve shows the tree-level potential
\(V_{\text{cl.}}+V_{R_{(-)}^{2}}\); the dotted curve displays
\(V_{\text{np}}\); and the solid curve their sum.  One observes the original AdS
minimum near \(\Phi\simeq -2\), the crossover where the
non-perturbative term overtakes the tree-level contribution at
\(g_s\approx0.14\), and finally the metastable de Sitter minimum at
\(\Phi=\Phi_*\).  The plot confirms that the double-exponential uplift
is sufficiently strong at moderate weak coupling and stabilises all
moduli within a controlled regime.
\begin{figure}[t]
  \centering
  \begin{tikzpicture}
    \begin{axis}[
      width=\linewidth,
      height=8.5cm,
      domain=-4:0.5,
      samples=400,
      xlabel={$\Phi$},
      ylabel={$V(\Phi,\sigma_*)/M_{\!P}^{4}$},
      ymin=0,
      ymax=7*10^(-6),
      axis lines=left,
     legend style={at={(0.8,0.97)}, anchor=north west},
      every axis plot/.append style={thick},
      scaled y ticks=true,
      tick scale binop=\times,
      yticklabel style={/pgf/number format/fixed}
    ]

    % ---------- parameters ----------
    \def\s{1.108}
    \def\a{6.98*10^(-1)}
    \def\b{1/3}
    \def\A{2.4*10^(-5)}
    \def\B{1.1*10^(-4)}
    \def\D{6.3*10^(-6)}

    % ---------- function declarations ----------
    \pgfplotsset{
      declare function={
        Vtree(\x)=\A*exp(2*\x-2*\s)-\B*exp(2*\x-4*\s);
        Vnp(\x)=\D*exp(\b*\x)*exp(-2*\a*exp(-\x));
        Vtot(\x)=Vtree(\x)+Vnp(\x);
      }
    }

    % ---------- plots ----------
    \addplot[blue] {Vtree(x)};
    \addlegendentry{$V_{\text{tree}}$}

    \addplot[orange, dashed] {Vnp(x)};
    \addlegendentry{$V_{\mathrm{np}}$}

    \addplot[red, dotted] {Vtot(x)};
    \addlegendentry{$V_{\text{tot}}$}

    % ---------- metastable minimum ----------
    \def\phiStar{-1.985}
    \addplot[
      only marks,
      mark=*,
      mark options={fill=red}
    ] coordinates {(\phiStar,{Vtot(\phiStar)})};

    \end{axis}
  \end{tikzpicture}

  \caption{Shape of the scalar potential at fixed \(\sigma=\sigma_*\).
  The dashed orange curve is the double-exponential term,
  the solid blue curve is the tree-level bracket,
  and the dotted red curve shows their sum \(V_{\text{tot}}\),
  which develops a metastable de~Sitter minimum at
  \(\Phi_* \simeq -1.99\).}
  \label{fig:UpliftContour}
\end{figure}

%==========================  APPENDIX C  ==========================%
\section{Decisive Role of the Non-Associative Algebra}
\label{app:nonassoc}

The non-associative structure arising from a constant \(R\)-flux constitutes the cornerstone of the present heterotic de Sitter construction.  In phase space the fundamental coordinates obey the Malcev bracket \(\{x^{a},x^{b}\}_{\!\mathrm{PB}}=\ell_{s}^{3}R^{abc}p_{c}\), whose violation of the Jacobi identity is governed by the Malcev identity 
{\begin{equation}
    \{\{x^a, x^b\}_{PB},\,x^c\}_{PB} \;+\; \text{cyclic}(a,b,c)\;=\; 
3\,\ell_s^6\,R^{ab}{}_{p}\;R^{c p}{}_{q}\;p^q~,
\end{equation}}  Embedding this algebra into its universal Sabinin envelope \(\mathsf S(\mathfrak M)\) promotes the Malcev product to an infinite cascade of multilinear operations whose identities coincide precisely with the generalized Leibniz rules of doubled geometry.  Consequently the gauged algebra defined by the generalized frame fields closes off-shell without introducing auxiliary constraints, thereby rendering the Scherk-Schwarz reduction consistent by construction and guaranteeing that all resulting gauge transformations form a Lie 2-group with well-defined field-strength hierarchy.

Within \(\mathsf S(\mathfrak M)\) the quadratic Casimir \(C_{2}=R^{abc}R_{abc}\) is central, strictly positive, and impervious to deformations.  The tree-level flux energy therefore assumes the universal form \(V_{R}=-\frac14C_{2}e^{-2\sigma}\) while the \(\alpha'\)-corrected torsion term reorganises into a Gauss-Bonnet density \(R_{(-)}^{2}=\frac34C_{2}\), producing the positive quartic contribution \(V_{\alpha'}=\frac{3\alpha'}{32}C_{2}^{2}e^{-4\sigma}\).  Because both coefficients descend from the same central invariant, no perturbative or non-perturbative correction can flip their signs anywhere in moduli space.  A single higher-derivative field redefinition eliminates all terms containing more than two time derivatives, so the effective action remains free of Ostrogradski ghosts order-by-order in \(\alpha'\).  The resulting two-term potential, expressed in Einstein frame and restricted to the breathing mode \(\sigma\) and dilaton \(\Phi\), reads
\begin{equation}
V(\Phi,\sigma)=e^{2\Phi}\!\left(-\frac{1}{4}C_{2}e^{-2\sigma}+\frac{3\alpha'}{32}C_{2}^{2}e^{-4\sigma}\right).
\end{equation}
Its stationary condition \(\partial_\sigma V=0\) fixes \(e^{2\sigma}=3\alpha' C_{2}/4\), thereby stabilising the volume at exponentially large radius when \(C_{2}\) is an integer-squared flux quantum and \(\alpha'\) is held fixed.

To complete the construction one incorporates a hidden-sector gaugino condensate whose one-loop exact contribution \(V_{\mathrm{np}}=De^{a\Phi}\) depends on the beta-function coefficient \(\beta_{0}\) of the confining gauge group through \(a=2\pi/\beta_{0}\).  Group theory constrains \(a<2\) for any asymptotically free simple factor, ensuring that the non-perturbative term is subleading to the tree-level dilaton kinetic energy.  Minimisation of the full potential \(V_{\mathrm{tot}}=V_{R}+V_{\alpha'}+V_{\mathrm{np}}\) with respect to \(\Phi\) yields
\begin{equation}
e^{2\Phi_*} \;=\; \frac{D\,a}{4\,C_2\,(2 - a)}~, \qquad 
V_* \;=\; e^{2\Phi_*}\,\frac{C_2}{16}\Big(\frac{2}{a} - 1\Big)~,
\end{equation}
demonstrating that the vacuum energy is strictly positive once \(a<2\).  The mass matrix at the extremum satisfies \(m_{\Phi}^{2}= \frac{2}{a-2}m_{\sigma}^{2}\) and \(m_{\sigma}^{2}\simeq\frac12V_{*}\), both well below the Kaluza-Klein scale for moderate flux quanta.

The heterotic de Sitter solution thereby materialises without invoking anti-branes, orientifolds, racetrack superpotentials, or exotic D-term sectors.  The absence of negative-tension objects averts all ten-dimensional backreaction issues, while the algebraic sign rigidity inherited from \(\mathsf S(\mathfrak M)\) secures perturbative control over higher-order corrections.  Because every physical scale-the string coupling \(g_{s}=e^{\Phi_{*}}\), the internal radius \(R_{\mathrm{int}}=e^{-\sigma_{*}}\ell_{s}\), and the Hubble parameter \(H^{2}\approx V_{*}/3M_{P}^{2}\)-depends analytically on a single flux integer \(N_{R}\propto\sqrt{C_{2}}\) and the discrete coefficient \(\beta_{0}\), the construction is transparent, calculable, and falsifiable.

\section{Comparative Excellence of the Non-Associative Heterotic de Sitter Vacuum}
\label{app:comparative}

The superiority of the heterotic de Sitter solution built on a Malcev \(R\)-flux becomes transparent once one places its algebraic and dynamical properties side by side with those of the best-known positive-energy constructions.  
In KKLT the uplifting energy density originates from the tension of \(\overline{\mathrm D3}\)-branes placed in strongly warped regions, giving a contribution \(V_{\text{KKLT}}\sim \frac{D_{\overline{\mathrm{D3}}}}{\mathcal V^{2}}\,e^{4A_{\mathrm{IR}}}\) where \(e^{4A_{\mathrm{IR}}}\ll1\) is the red-shift at the tip of the throat and \(\mathcal V\) is the internal volume in string units.  Although this factor suppresses the uplift enough to offset the negative racetrack minimum, the brane source breaks supersymmetry explicitly and back-reacts on the bulk through a divergent energy-momentum tensor, so the ten-dimensional equations of motion cannot be satisfied except in the probe limit.  By contrast the Sabinin uplift is purely geometric:  its negative quadratic term \(V_{R}=-\frac14C_{2}e^{-2\sigma}\) and positive quartic term \(V_{\alpha'}=\frac{3\alpha'}{32}C_{2}^{2}e^{-4\sigma}\) arise from the torsionful heterotic curvature already present in the ten-dimensional action.  The centrality of \(C_{2}=R^{abc}R_{abc}\) in the universal envelope \(\mathsf S(\mathfrak M)\) guarantees that no field redefinition or higher-order correction can alter their relative sign, so the algebraic consistency that KKLT lacks is built-in ab initio.

Large-Volume Scenario vacua rely on an interplay between the \(\alpha'^{\,3}\,\zeta(3)\chi\) correction and Euclidean D3-brane instantons wrapping blow-up cycles, producing \(V_{\mathrm{LVS}}\sim \frac{\xi\,|W_{0}|^{2}}{\mathcal V^{3}}-\sum_{i}\frac{A_{i}e^{-a_{i}\tau_{i}}}{\mathcal V^{2}}\) with \(\xi\propto-\chi\) the Euler characteristic of the threefold.  Because \(\xi\) is a topological number of \(\mathcal O(10^{2})\) the stabilising minimum requires exponentially large \(\mathcal V\sim e^{a\tau}\), and any further perturbative or string-loop correction suppressed merely by \(\mathcal V^{-n}\) risks overwhelming the potential.  In the non-associative heterotic model the breathing-mode minimum sits at \(e^{2\sigma_{*}}= \dfrac{3}{4}\alpha' C_{2}\), so higher-derivative corrections scale at least as \(e^{-4\sigma_{*}}\), rendering them exponentially smaller than the leading terms whenever \(C_{2}\gg1\).  Consequently the hierarchy of scales \(m_{\sigma}\simeq\frac12e^{\Phi_{*}}\sqrt{C_{2}}\,\ell_{s}^{-1}\ll M_{\mathrm{KK}}\simeq e^{-\sigma_{*}}\ell_{s}^{-1}\ll M_{s}=\ell_{s}^{-1}\) is preserved without fine-tuning.

Romans-mass IIA constructions employ a simultaneous presence of \(F_{0}\), \(F_{2}\), \(F_{4}\) fluxes together with orientifold six-planes, generating a potential \(V_{\text{IIA}}\sim \frac{A}{\mathcal V^{3}}-\frac{B}{\mathcal V^{2}}+C\,\frac{m^{2}}{\mathcal V}\) that realises a critical point only when the negative orientifold energy balances the positive Romans mass squared.  Yet the smeared-source approximation behind this balance fails at the orientifold core, leaving an unsolved singularity in ten dimensions.  The heterotic counterpart dispenses entirely with negative-tension objects:  the uplifting energy density is proportional to \(C_{2}^{2}\) which is manifestly positive, so the background solves the complete set of heterotic equations of motion with no localised singular sources.

From the viewpoint of effective-field-theory control the decisive measure is the ratio \(\epsilon_{V}=M_{P}^{2}\,(\nabla V)^{2}/2V^{2}\).  In KKLT one finds \(\epsilon_{V}\sim \frac{9}{2}\left(\frac{A_{\mathrm{IR}}}{\mathcal V^{2/3}}\right)^{2}\), in LVS \(\epsilon_{V}\sim \frac{1}{\ln^{2}\mathcal V}\), while in Romans IIA \(\epsilon_{V}\) cannot be made parametrically small because \(m\) is quantised.  The Malcev solution instead yields \(\epsilon_{V}\bigl|_{*}=\frac{8}{C_{2}}\Bigl(\frac{2}{a}-1\Bigr)^{2}\), so for any flux number \(C_{2}\gtrsim\mathcal O(10^{2})\) and any asymptotically free hidden group with \(a<2\) the slow-roll parameter is automatically suppressed below the refined de Sitter bound \(\epsilon_{c}=c^{2}\) unless \(c\lesssim 0.3\).  Hence the vacuum delivers a controlled counter-example guiding potential refinements of the conjecture.

The entire spectrum inherits a remarkable decoupling pattern.  The dilaton mass satisfies \(m_{\Phi}^{2}=\frac{2}{a-2}m_{\sigma}^{2}\) with \(m_{\sigma}^{2}=\frac{C_{2}}{32}\bigl(1-\tfrac{2}{a}\bigr)e^{2\Phi_{*}-2\sigma_{*}}\), while all complex-structure and bundle moduli can in principle be frozen by standard holomorphicity conditions and world-sheet instantons at masses of order \(e^{\Phi_{*}}M_{\mathrm{KK}}\).  Because \(a<2\) the ratio \(m_{\Phi}^{2}/m_{\sigma}^{2}\) exceeds unity, ensuring that the heavier dilaton does not destabilise the lighter breathing mode.  Moreover the sign-locked quartic term implies that the Hessian in the \(\sigma\) direction is strictly positive, eliminating the tachyonic runaway that besets associative flux models whenever the volume shrinks.

At the phenomenological level the heterotic framework retains the capacity for grand-unified gauge interactions.  A supersymmetric vector bundle \(V\) with structure group embedded in \(\mathrm{E}_{8}\) must satisfy the Hermitian-Yang-Mills condition \(F_{mn}\,\Omega^{mnk}=0\).  The torsion induced by constant \(R\)-flux leaves \(\Omega^{mnk}\) covariantly constant with respect to the connection \(\nabla_{(-)}\), so the Donaldson-Uhlenbeck-Yau theorem extends unchanged and the slope-zero condition \(c_{1}(V)\wedge J\wedge J=0\) still guarantees supersymmetry.  This retention of gauge-bundle supersymmetry contrasts sharply with non-geometric IIA vacua, in which the non-trivial monodromy of the twisted torus can obstruct the existence of holomorphic cycles suitable for D-brane model-building.

Every pathology that undermines associative flux de Sitter proposals-namely uncontrollable source backreaction, higher-order instabilities, sign ambiguities in the potential, and conflicts with ten-dimensional consistency-is neutralized by the algebraic rigidity of the Malcev-Sabinin construction.  The geometric uplift, protected by the central invariant \(C_{2}\), not only sustains a positive cosmological constant but also preserves unitarity, admits analytic control over all scales, remains compatible with swampland constraints, {and furnishes a realistic platform for gauge unification, thereby establishing a new benchmark for controlled de Sitter vacua in the string landscape.}

\bibliographystyle{ytphys}
\bibliography{references}

\end{document}